\documentclass[preprint,aps,floats,nofootinbib,showpacs]{revtex4}
\usepackage{epsfig,epsf}
\usepackage{amssymb}
\usepackage{amsmath}
\usepackage{graphicx}
\newcommand{\be}{\begin{equation}}
\newcommand{\ee}{\end{equation}}
\newcommand{\gsim}{\, \raisebox{-0.8ex}{$\stackrel{\textstyle >}{\sim}$ }}
\newcommand{\lsim}{\, \, \raisebox{-0.8ex}{$\stackrel{\textstyle <}{\sim}$ }}

\newcommand\imag{ i }

\newcommand{\roughly}[1]%
{\mathrel{\raise.4ex\hbox{$#1$\kern-.75em\lower1ex\hbox{$\sim$}}}}

\newcommand\CE{{\cal E}}
\newcommand\CL{{\cal L}}

\newcommand\beq{\begin{eqnarray}}
\newcommand\eeq{\end{eqnarray}}
\newcommand\eqn[1]{\label{eq:#1}}

\def\Dsl{\,\raise.15ex \hbox{/}\mkern-12.8mu D}
\newcommand\Tr{{\rm Tr\,}}

\def\fm3{fm$^{-3}$}

\def\ni{\noindent}
\begin{document}
%
\preprint{\vbox{\hbox{LA-U R- ?????}}}

\bigskip
\bigskip

\title{Neutrino processes in the $K^0$ condensed phase of color flavor
locked quark matter }

\author{Sanjay Reddy$^1$, Mariusz Sadzikowski$^{2}$, Motoi Tachibana$^3$}

\affiliation{Theoretical Division, Los Alamos National Laboratory, Los
Alamos, NM 87545 \\ $^2$M. Smoluchowski Institute of Physics,
Jagellonian University, Reymonta 4, 30-059 Krak\'ow, Poland \\ $^3$
Theoretical Physics Laboratory, RIKEN, Wako 351-0198, Japan. }
\begin{abstract}
We study weak interactions involving Goldstone bosons in the neutral
kaon condensed phase of color flavor locked quark matter. We calculate
the rates for the dominant processes that contribute to the neutrino mean free path and to neutrino production. A light $K^+$ state, with a mass
$\tilde{m}_{K^+} \propto (\Delta/\mu)~(\Delta/m_s)(m_d-m_u)$, where
$\mu$ and $\Delta$ are the quark chemical potential and
superconducting gap respectively, is shown to play an important role.
We  identify unique characteristics of weak interaction rates in this
novel phase and discuss how they might influence neutrino emission in
core collapse supernova and neutron stars.
\end{abstract}
\maketitle

\section{Introduction}
At high baryon density, where the baryon chemical potential
$\mu$ is very large compared to the strange quark mass $m_s$, three
flavor quark matter is expected to be in a symmetric phase called the
Color Flavor Locked (CFL) phase, in which BCS like pairing involves
all nine quarks \cite{Alford:1998mk}. This, color superconducting,
phase is characterized by a gap in the quark excitation
spectrum. Model calculations indicate that the gap $\Delta \sim 100$
MeV for a quark chemical potential $\mu \sim 500$ MeV
\cite{Alford:1998zt,Rapp:1998zu}.  In this phase, the ${ SU(3)_{\rm
color}} \times SU(3)_L \times SU(3)_R \times U(1)_B$ symmetry of QCD
is broken down to the global diagonal $SU(3)$ symmetry. The lightest
excitations in this phase are the nonet of pseudo-Goldstone bosons
transforming under the unbroken, global diagonal $SU(3)$ as an octet
plus a singlet and a massless mode associated with the breaking of the
global $U(1)_B$ symmetry. At lower density, those that are of
relevance to neutron stars, the baryon chemical potential $\mu$ is
similar to $m_s$. At these densities, the strange quark mass induces a
stress on the SU(3) flavor symmetric CFL state. Bedaque and Schafer
\cite{Bedaque:2001je} have shown that the stress induced by the
strange quark mass can result in the condensation of neutral kaons in
the CFL state. This less symmetric phase, which is called the CFLK$^0$
phase, breaks hypercharge $U(1)_Y$ and isospin symmetries and we can expect its low energy properties to be quite distinct from the CFL phase. For a detailed discussion of various meson condensed phases in CFL quark
matter and their possible role in compact stars see
Ref.~\cite{Kaplan:2001qk}.

In this article we extend earlier work (see
Refs.~\cite{Reddy:2002xc,Jaikumar:2002vg}) on neutrino emission and
propagation rates in the CFL phase to the CFLK$^0$ phase. As discussed
in detail in these aforementioned references, novel phases inside
compact stars, if they should exist, influence the stars thermal
evolution. In particular, neutrino production and propagation rates in
these novel phases can directly influence observable aspects such as
the supernova neutrino emission rate and longer term neutron star
cooling rates \cite{Prakash:2000jr}.

\section{Thermodynamics of the CFLK$^0$ Phase}

The low energy excitations about the $SU(3)$ symmetric CFL ground
state can be written in terms of the two fields: $B=H /(\sqrt{24}f_H)$
and $\Sigma=e^{2i({\pi}/f_\pi+\eta'/f_A)}$, representing the Goldstone
bosons of broken baryon number $H$, and of broken chiral symmetry, the
pseudo-scalar octet ${ \pi}$, and the pseudo-Goldstone boson $\eta'$,
arising from broken approximate $U(1)_A$ symmetry.  The leading terms
of the effective Lagrangian describing the octet Goldstone boson field
$\pi$ is given by
\beq \eqn{leff} \CL &=&
\frac{1}{4}f_{\pi}^2 \left[\Tr \nabla_0 \Sigma \nabla_0 \Sigma^\dagger
- v^2 \Tr \vec{\nabla}\Sigma \cdot \vec{\nabla} \Sigma^\dagger \right]
\ \, \nonumber \\ && + f_{\pi}^2 \left[\frac{a}{2} \Tr \tilde{M}
\left(\Sigma + \Sigma^\dagger\right) + \frac{\chi}{2} \Tr {M}
\left(\Sigma +\Sigma^\dagger\right)\right] \ \,, \nonumber \\ &{\rm where} &
\nabla_0\Sigma = \partial_0 \Sigma - i\left[ (\mu_Q  Q - X_L) \Sigma -
\Sigma (\mu_Q Q - X_R) \right] \,.
\label{eqn:oldlagrangian}
\eeq
The decay constant $f_\pi = 0.21 \mu $ has been computed previously
\cite{Son:2000cm} and is proportional to the quark chemical
potential. The quark mass matrix $M={\rm diag}  (m_u,m_d,m_s)$.
 $X_{L,R}$ are the Bedaque-Schafer terms: $X_L = \frac{M
M^\dagger}{2\mu}\ ,\quad X_R = \frac{ M^\dagger M}{2\mu}\ \,, $
$\tilde{M}= |M|M^{-1}$ and $\mu_Q$ is the electric charge chemical
potential associated with the unbroken $U(1)$ in the CFL phase . A
finite baryon chemical potential breaks Lorentz invariance of the
effective theory. The temporal and spatial decay constants can thereby
differ. This difference is encoded in the velocity factor $v$ being
different from unity. An explicit calculation shows that
$v=1/\sqrt{3}$ and is common to all Goldstone bosons, including the
massless $U(1)_B$ Goldstone boson\cite{Son:2000cm}. At asymptotic
densities, where the instanton induced interactions are highly
suppressed and the $U(1)_A$ symmetry is restored, the leading
contributions to meson masses arise from the Tr$ \tilde{M} \Sigma$
operator whose coefficient $a$ has been computed and is given by
$a=3\frac{\Delta^2}{\pi^2 f_{\pi}^2}$
\cite{Son:2000cm,Schafer:2001za}. At densities of relevance to neutron
stars the instanton interaction may become relevant. In this case a
$\langle \bar{q} q\rangle$ condensate is induced
\cite{Manuel:2000wm,Schafer:2002ty}. Consequently, the meson mass can
receive a contribution from the operator Tr$ M \Sigma$.  Its
coefficient, $\chi$, at low density is poorly known and is sensitive
to the instanton size distribution and form factors. Although,
conservative current estimates indicates that the instanton
contribution to the $K^0$ mass for $\mu\sim 400$ MeV lies in the range
$5-120$ MeV theory favors a value that is $\lsim 30 $ MeV
\cite{Schafer:2002ty}.  The $\eta'$ mass is also poorly known at
moderate density. The contribution to the $\eta'$ mass due to the
$U(1)_A$ anomaly, which does {\it not} vanish in the chiral limit, is not known. If this contribution is negligible, the neutral $\pi^0$, $\eta$ and
$\eta'$ mix due to the explicit breaking of SU(3)$_{\rm flavor}$ by
the quark masses. In our investigation here we will assume that the
anomalous contribution to the $\eta'$ mass is large and that the
$\eta'$ decouples from the low energy theory. We will return to
comment on how a light $\eta'$ will affect our results in \S \ref{caveats}.

The meson masses, written in terms of the coefficients a and $\chi$
are given by
\beq
m^2_{\pi^\pm} &=& a (m_u + m_d)m_s + \chi (m_u+m_d) \cr
m^2_{K^\pm} &=& a (m_u + m_s)m_d + \chi (m_u+m_s)  \cr
m^2_{K^0} &=& a (m_d +m_s)m_u + \chi (m_d+m_s)\,.
\label{masses1}
\eeq
The dispersion relations for Goldstone modes in the CFL phase are
unusual, as will become clear from the following discussion.  They are
easily computed by expanding the effective Lagrangian to second order
in meson fields
\beq
E_{\pi^\pm}(p) &=& \sqrt{v^2 p^2+ m_{\pi_\pm}^2} \quad
E_{K^+}(p) = -X + \sqrt{v^2 p^2+ m_{K^+}^2} \nonumber \\
E_{K^-}(p) &=& X + \sqrt{v^2 p^2+ m_{K^-}^2} \quad
E_{K^0}(p) = -X + \sqrt{v^2 p^2+ m_{K^0}^2} \,,
\label{masses2}
\eeq
where $X=m_s^2/2\mu$.  We note that in deriving the above relations we
have neglected terms of order $m^2_{\rm light}/\mu$ since they are
negligible compared to all other relevant scales in the problem,
namely $\mu,~m_s$ and $\Delta$. Meson dispersion relations violate
Lorentz invariance and the induced effective chemical potential
arising from the analysis of Bedaque and Schafer \cite{Bedaque:2001je}
breaks the energy degeneracy of the kaons and anti-kaons. In
Ref.~\cite{Reddy:2002xc} we had chosen the instanton contribution to
the $K^0$ mass is $\sim 50$ MeV (corresponding to $\chi \sim 15 $ MeV)
at $\mu=400$ MeV and $\Delta=100$ MeV.  For this choice the
kaon mass is too large to allow for $K^0$ condensation. If the
instanton contribution is small at $\mu\sim400$ MeV, the CFL phase
becomes unstable to $K^0$ condensation when $E_{K^0}(p=0)<0$ or $X \ge m_{K^0}$\cite{Bedaque:2001je}. In this article, wherein we investigate
weak interaction processes in the $K^0$ condensed phase we assume that
the instanton contribution is negligible at $\mu=400$ MeV and set
$\chi=0$.  Furthermore, the numerical results we present in subsequent
sections will be for the specific choice of $\mu=400$ MeV, $\Delta=100$ MeV, $m_u=3.75$ MeV, $m_d=7.5$ MeV and $m_s=150$ MeV.

In the CFLK$^0$ phase the ground state expectation value of
$\Sigma=\Sigma_0 \ne \bf{1}$ (in the symmetric CFL phase the
expectation value of $\Sigma=\bf{1}$). As discussed in
Ref.~\cite{Kaplan:2001qk}, the CFLK$^0$ phase is characterized by
\be
\Sigma_0=\left[ \begin{array}{ccc}
 1 & 0 & 0 \\
0& \cos{\theta}&  \imag \, \sin{\theta } \\
 0 & \imag \,\sin{\theta} & \cos{\theta} \\
\end{array} \right] \,,
\ee
where $\cos{\theta} = m_{K^0}^2/X^2$ and the number density of
condensate $n_{K^0}=f_{\pi}^2 X \sin^2{\theta}$.

To study the spectrum of low energy excitations in this phase we
perturb about this new ground state \cite{Schafer:2001bq}.  We can
incorporate the effect of the background kaon condensate by replacing
$\Sigma=\exp (2i \pi /f_\pi) $ by $\tilde{\Sigma}=\xi \Sigma \xi$ in
Eq. \ref{eqn:oldlagrangian}
\footnote{This simple parameterization was originally suggested by
David Kaplan, see also Ref. \cite{Kryjevski:2002ju}.}, where
$\xi=\sqrt{\Sigma_0}$. The matrix $\pi=\pi_a T_a$ where $T_{a=1..8}$
are the generators of SU(3) and $\pi_{a=1..8}$ are real scalar fields
characterizing the excitations about the CFLK$^0$ phase.  Replacing
$\Sigma \rightarrow \tilde{\Sigma}$ in Eq. \ref{eqn:oldlagrangian} and
expanding to quadratic order in the fields $\pi_{a=1..8}$, we obtain the
leading order Lagrangian
\beq
\mathcal{L} &=& \mathrm{Tr} [ \partial_t  \pi  \partial_t \pi ]
- v^2 \mathrm{Tr}  [\nabla \pi \cdot \nabla \pi ]
- \imag \mathrm{Tr} [\partial_t \pi ([\mu_r,\pi]+[\mu_l,\pi])] \nonumber \\
&-&\mathrm{Tr} [[\mu_r,\pi]~[\mu_l,\pi]]
- a \mathrm{Tr}[(\tilde{M_l}+\tilde{M_r})\pi^2]
- \chi \mathrm{Tr}[(M_l+M_r)\pi^2]
\label{eqn:newlagrangian}\, .
\eeq
where
\begin{alignat}{2}
\mu_r &=\xi ~(\mu_Q Q + X_R)~ \xi^\dagger   & \qquad
\mu_l &=\xi^\dagger~ (\mu_Q Q + X_L)~ \xi    \\
\tilde{M_r}&=\xi^\dagger ~\mathrm{det} M ~M^{-1} ~\xi^\dagger     &
\tilde{M_l}&=\xi ~\mathrm{det} M~ M^{-1}~ \xi   \\
M_r & = \xi^\dagger ~M ~\xi^\dagger    &   M_l &=\xi~ M ~\xi \,.
\end{alignat}
The equations of motion for the fields $\pi_{a=1\cdots8}$ are given by
\beq
(\omega^2 - v^2 k^2) \pi_a &=& \kappa_{ab}~\pi_b  \nonumber \\
\kappa_{ab}&=&\mathrm{Tr}[[\mu_l,t_a][\mu_r,t_b]+[\mu_l,t_b][\mu_r,t_a]]
\nonumber \\ &+&\omega~\mathrm{Tr}[t_b[(\mu_l+\mu_r),t_a] -
t_a[(\mu_l+\mu_r),t_b]] \nonumber \\ &+&
a~\mathrm{Tr}[(\tilde{M_l}+\tilde{M_r})(t_a t_b + t_b t_a)] \nonumber
\\ 
&+& \chi~\mathrm{Tr}[(M_l+M_r)(t_a t_b + t_b t_a)]\,,
\label{mastereqn}
\eeq
where $\omega,k$ are the energy and momenta of the propagating modes.

Mixing arising due to non-diagonal components of $\kappa_{ab}$ is
sparse. Spontaneous breaking of $U(1)_Y$ in the $K^0$ condensed phase
leads to strong mixing between $K^0$ and $\bar{K}^0$ states.
The dispersion relations for these neutral kaons are given by
\beq
\omega_{K_1}^2&=&v^2p^2+\frac{X^2}{2}\left[(1+3\cos^2{\theta})
-\sqrt{(1+3\cos^2{\theta})^2
+ 16 \cos^2{\theta} \frac{ v^2p^2}{X^2}}\right] \\
\omega_{K_2}^2&=&v^2p^2+\frac{X^2}{2}\left[(1+3\cos^2{\theta})
+\sqrt{(1+3\cos^2{\theta})^2
+ 16 \cos^2{\theta} \frac{ v^2p^2}{X^2}}\right] \,.  \eeq
For $v~ p \ll X$ they are given by
\beq \omega_{K_1}^2&=&
\frac{1-\cos^2{\theta}}{1+3\cos^2{\theta}} ~ v^2p^2 + \frac{16
\cos{\theta}^4}{(1+ 3 \cos^2{\theta})^3}~\frac{v^4p^2}{X^2} +
\mathcal{O}\left[\frac{v^6p^6}{X^4}\right] \\
\omega_{K_2}^2&=&(1+3\cos^2{\theta})X^2 +
\frac{1+7\cos^2{\theta}}{1+3\cos^2{\theta}}~ v^2p^2 + \frac{16
\cos{\theta}^4}{(1+ 3 \cos^2{\theta})^3}~ \frac{v^4p^2}{X^2} +
\mathcal{O}\left[\frac{v^6p^6}{X^4}\right]
\eeq
The $K_1$ mode is the massless Goldstone boson which we expected on
general grounds since the ground state breaks hypercharge symmetry. As an aside, we note that when iso-spin is not explicitly broken by the quark masses, the $K^+$ is also massless mode  and has a quadratic dispersion
relation $\omega_{K^+} = v^2 p^2/2Y + (v^4 p^4) $. This latter mode, with its quadratic dispersion relation, has been shown to account for two broken generators and hence alters the number of expected Goldstone modes \cite{Schafer:2001bq,Miransky:2001tw}.   

In contrast, since $U(1)_Q$ remains unbroken the charged kaon (and
pion) states do not mix.  The dispersion relations for the charged
kaons and pions are given by
\beq
\omega_{K^{\pm}} &=& \mp (Y +\mu_Q) + \sqrt{v^2p^2 + Y^2 + \delta m^2} \\
\omega_{\pi^{\pm}}&=& \mp (Z +\mu_Q) + \sqrt{v^2 p^2 + Z^2 + m_{\pi^+}^2}
\,,
\label{chargedmodes}
\eeq
\begin{figure}[]
\begin{center}
\includegraphics[width=.48\textwidth,height=0.48\textwidth]{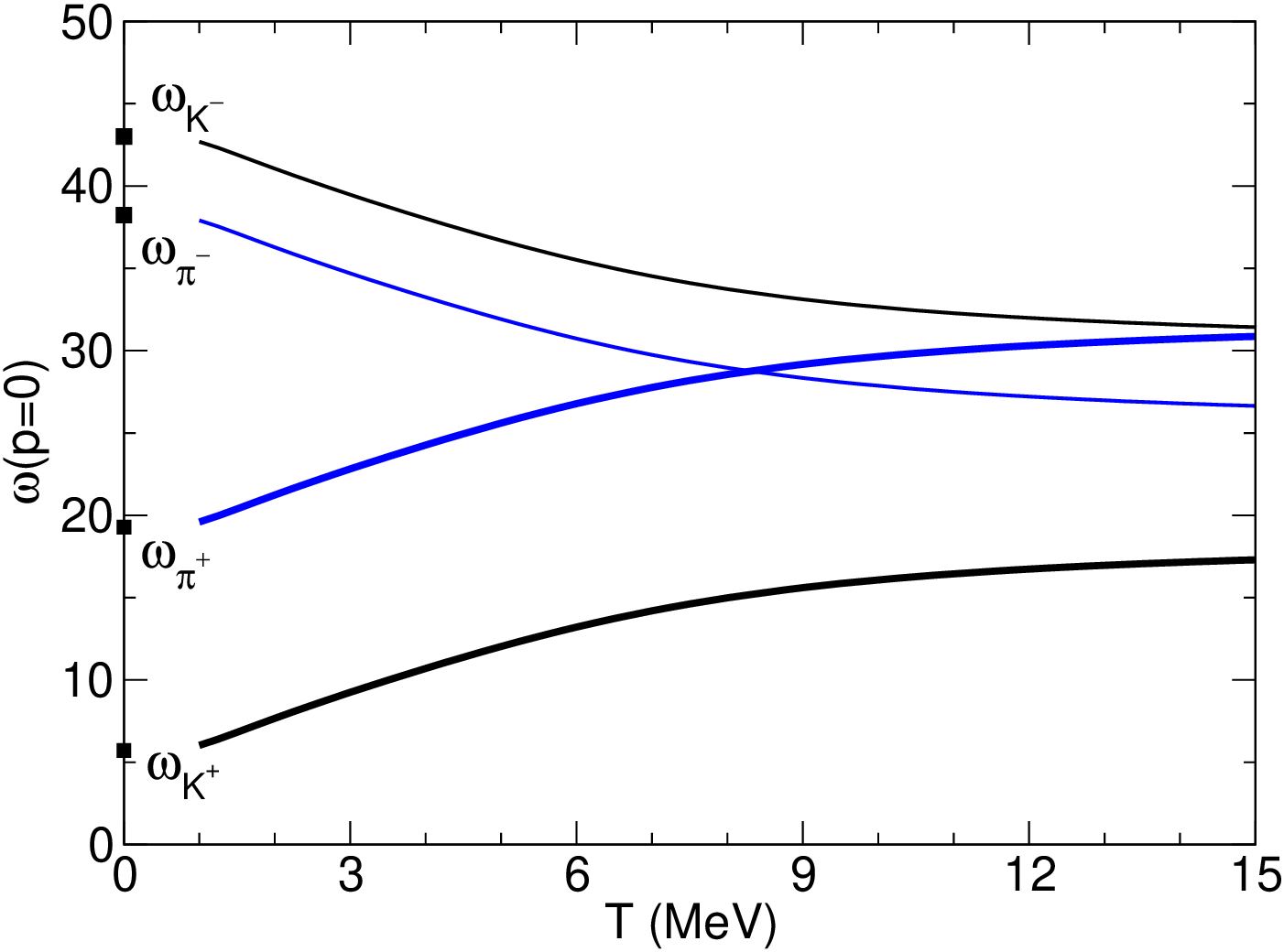}
\includegraphics[width=.48\textwidth,height=0.48\textwidth]{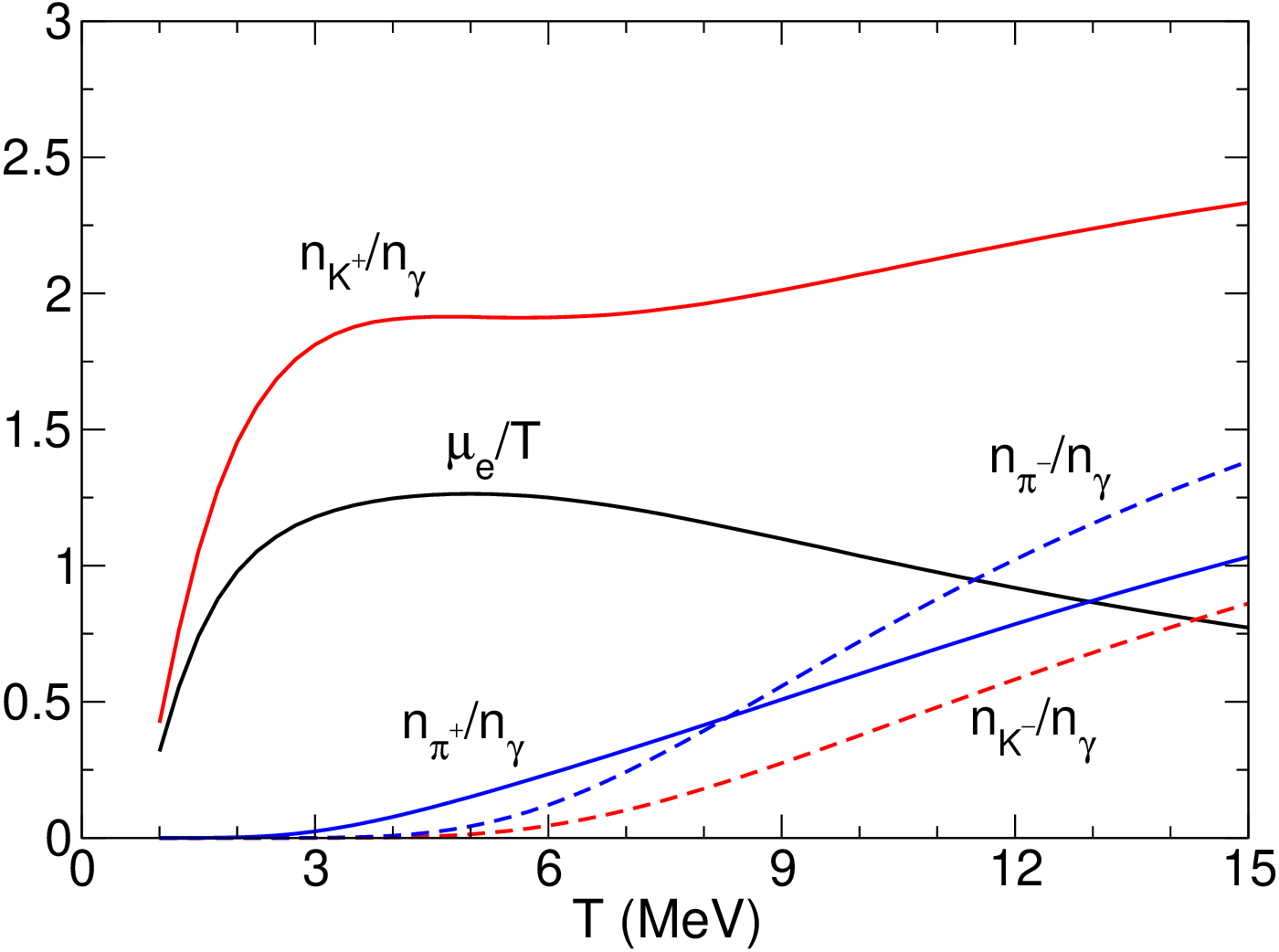}
\caption{Left panel: The excitation energy (see
Eq.~\ref{chargedmodes}) of the charged Goldstone modes in the $K^0$
phase. The variation with temperature arises because of the induced
charge chemical potential. The square symbols on the y-axis correspond
to the masses at $T=0$. Right Panel: Charged particle densities
normalized by the photon number density and the induced electric
charge chemical potential normalized by the temperature in the
CFLK$^0$ phase.}
\label{number}
\end{center}
\end{figure}
where $Y=X(1+\cos{\theta})/2$, $ Z=X(1-\cos{\theta})/2$ and $\delta
m^2=m_{K^+}^2-m_{K^0}^2$. The mass of the $K^+$ mode in the CFLK$^0$ phase $\tilde{m}_{K^+} = \sqrt{Y^2 + \delta m^2} - Y$ and is the
lightest charged excitation. When the condensate amplitude is large,
i.e. $\cos{\theta}\simeq 0$, the $\pi^+$ is also a relatively
light excitation with a mass $\tilde{m}_{\pi^+}=\sqrt{Z^2 + m_{\pi}^2}
- Z$.

As discussed in Ref.~\cite{Reddy:2002xc}, differences in the
excitation energies of mesons with positive and negative charge will
result in a net charge density in the meson gas at finite
temperature. With increasing temperature a finite electric charge
chemical potential is induced to ensure that matter is electrically
neutral. The excess positive electric charge of the meson gas is
compensated by electrons with chemical potential $\mu_e=-\mu_Q$.
Fig.~\ref{number} shows the excitation energy and the number density of
charged mesons (normalized by the photon number density
$n_\gamma=2\zeta(3) T^3/\pi^2$) and the electron chemical potential
$\mu_e$ (normalized by the temperature) as a function of the ambient
temperature. The two massless modes, namely the $H$ and $K_1$ are the
most abundant species. The ratio $n_H/n_\gamma=1/(2v^3)\simeq 2.6$, and for $T\ll X$ the ratio $n_{K_1}/n_\gamma=1/v_{K_1}^3$ where
$v_{K_1}=\sqrt{(1-\cos^2{\theta})/(1+3\cos^2{\theta})}~v$ is the
velocity of the $K_1$ mode. For $T\ll X $ and for the numerical values
of the parameters chosen, we find that $n_{K_1}/n_\gamma \simeq 4$.
This indicates that the thermodynamic properties of the CFLK$^0$ phase
is dominated by the H and $K_1$ modes. The low temperature specific
heat of the $K^0$ phase is roughly twice as large as the CFL phase.

On general grounds we can expect the amplitude of the kaon condensate
to decrease with increasing temperature. A precise calculation of the
critical temperature at which the condensate melts is beyond the
scope of this work. However, we can make a rough estimate by noting
that $X=m_s^2/2\mu$ is the important dimensional scale that
characterizes the condensate at zero temperature (we assume a robust
condensate that $X \gg m_{k^0}$) . A weakly interacting Bose Einstein
condensate melts when the thermal wavelength becomes comparable to
the inter-particle distance in the ground state. The inter-particle
distance in the ground state $l=(1/n_{K^0})^{1/3}$ where $n_{K^0}
\simeq f^2_{\pi}X$. Numerical values of the parameters employed, we can expect $T_c \gsim 50$ MeV. For $T$ small compared to $T_c$, it is reasonable to assume that finite temperature effects on the propagation of Goldstone modes in the CFLK$^0$ phase can be ignored. Further, we note that corrections due to thermal loops are suppressed by the factor $T/f_\pi$ due to the derivative coupling between the mesons.  In the subsequent discussion of weak interaction rates we will restrict ourselves to these low
temperatures where $T \ll T_c \lsim f_\pi$.

\section{Neutrino Processes}
We now turn to the calculation of weak interaction rates in the
CFLK$^0$ phase. To compute the weak couplings of  Goldstone modes
in the $K^0$ phase, which we represented by a $3 \times 3$ matrix
$\Phi$, we gauge the chiral Lagrangian in the presence of the
background field characterized by $\xi$. We replace the derivative term of the original Lagrangian by the covariant derivative and this generates the leading order couplings. The gauged chiral Lagrangian is given by
\begin{equation}
\label{L}
{\cal L} = \frac{f_{\pi }^2 }{4}~Tr\left[ D_\rho\Sigma_\Phi
(D^\rho\Sigma_\Phi )^\dagger\right] \,, \\
\end{equation}
where the covariant derivative
\begin{eqnarray}
D_\rho\Sigma_\Phi &=& \nabla_\rho\Sigma_\Phi -
\frac{ig}{\sqrt{2}}(W^{+}_\rho\tau^+ + W^{-}_\rho\tau^-)\Sigma_\Phi
- \frac{\imag g~Z_\rho}{\cos\theta_W} (\tau^W_3\Sigma_\Phi - \sin\theta_W
[Q,\Sigma_\Phi]) - \imag \tilde{e}A_\rho [Q,\Sigma_\Phi] \,, \nonumber \\
\nabla \Sigma_\Phi &=& (\partial_0 \Sigma_\Phi + i~ \left[X_L\Sigma_\Phi
- i\Sigma_\Phi X_R\right],~\vec{\partial}\Sigma_\Phi )\,,\quad {\rm and}
\nonumber \\
\Sigma_\Phi&=&\xi~\exp{\left(\frac{2~\imag~\Phi}{f_\pi}\right)}~\xi \,.
\label{gauged}
\end{eqnarray}
In the above equation, $\theta_W$ is weak mixing angle, $Q$ is the
quark charge matrix, $\tau^W_3=1/2~{\rm diag}~(1,-1,-1)$ is the weak
isospin matrix and $\tau^\pm$ are usual charged current raising and
lower operators which include the Cabibo mixing and are given by
\be
\tau^+=\left[ \begin{array}{ccc}
0 & \cos{\alpha} & \sin{\alpha} \\
0 &  0 & 0 \\
0 & 0 & 0 \\
\end{array} \right] \,,
\qquad \quad
\tau^-=\left[ \begin{array}{ccc}
 0 & 0 & 0 \\
\cos{\alpha} &  0 & 0 \\
\sin{\alpha} & 0 & 0 \\
\end{array} \right] \,.
\ee
As discussed earlier the factor $v=1/\sqrt{3}$ breaks Lorentz invariance
and Minkowski  product is given by
\begin{equation}
D_\rho\Sigma_\Phi (D^\rho\Sigma_\Phi )^\dagger = D_0\Sigma_\Phi
(D^0\Sigma_\Phi )^\dagger
- v^2 \vec{D}\Sigma_\Phi (\vec{D}\Sigma_\Phi )^\dagger \,.
\end{equation}
Expanding the kinetic term in inverse powers of $f_\pi $ we can
identify the leading order couplings of the Goldstone modes to the
neutral and charged weak currents. At order $f_\pi$, the charged current weak interactions is described by
\begin{eqnarray}
\label{l1_cc}
{\cal L}^{(1)}_{cc} &=& -\frac{g f_\pi }{\sqrt{2}}~Tr\left[
\nabla^R \Phi ~\left[ \tilde{\tau}^+~W^+ +
\tilde{\tau}^-~W^-\right] \right] \,, \\
{\rm where}&& \nonumber \\
\nabla_{\sigma }^R\Phi &=& (\partial_0 \Phi - i[\Phi ,\xi X_R\xi^\dagger
], \vec{\partial }\Phi )\,, \\
\tilde{\tau}^+ &=& \xi^\dagger~\tau^+~\xi \qquad \tilde{\tau}^-=\xi^\dagger~\tau^-~\xi \,.
\end{eqnarray}
We note that the condensate, which introduces mixing between down and
strange quarks, modifies the Cabibo suppression of charged current
reactions. The {\it rotated} raising and lowering operators
$\tilde{\tau}_\pm$ are given by
\be
\tilde{\tau}^+=\left[ \begin{array}{ccc}
0 & \cos{\tilde{\alpha}}^\ast & \sin{\tilde{\alpha}}^\ast \\
0 &  0 & 0 \\
0 & 0 & 0 \\
\end{array} \right] \,,
\qquad
\tilde{\tau}^-=\left[ \begin{array}{ccc}
 0 & 0 & 0 \\
\cos{\tilde{\alpha}} &  0 & 0 \\
\sin{\tilde{\alpha}} & 0 & 0 \\
\end{array} \right] \,,
\ee
where
\begin{eqnarray}
\sin{\tilde{\alpha}} &=&\cos{\frac{\theta}{2}} \sin{\alpha}
- \imag~ \sin{\frac{\theta}{2}} \cos{\alpha}\,, \\
\cos{\tilde{\alpha}} &=&\cos{\frac{\theta}{2}} \cos{\alpha}
- \imag~ \sin{\frac{\theta}{2}} \sin{\alpha}\,.
\label{newcabibo}
\end{eqnarray}
Note, although $\tilde{\alpha}$ is complex, $|\sin{\tilde{\alpha}}|^2
+ |\cos{\tilde{\alpha}}|^2=1$.

The Lagrangian describing neutral current interactions at leading
order can be obtained similarly from Eq.~\ref{gauged}. Here, since
$\xi$ commutes with both $\tau^W_3$ and $Q$, we find that
\begin{eqnarray}
\label{l1_nc}
{\cal L}^{(1)}_{nc} = -\frac{g f_\pi }{\cos\Theta_W}~ Z^\sigma
~Tr\left[\nabla{\sigma }^R\Phi~\tau^W_3 \right] \,.
\end{eqnarray}
\subsection{Neutrino Opacity}
Two classes of processes involving mesons contribute to the neutrino
opacity.  They are $\nu\rightarrow e\phi $ and $\nu\phi\rightarrow e$
where $\phi = K^\pm ,\pi^\pm $.  These processes are kinematically
forbidden in the vacuum since meson dispersion relations are
restricted to be time-like with $\omega(p) > p$. In the medium, where
Lorentz symmetry is broken, mesons on mass-shell can acquire a
space-like dispersion relation.  This allows for novel processes
wherein a relativistic neutrino can radiate or absorb mesons
\cite{Reddy:2002xc}. These processes dominate the opacity since they
are appear at leading order and are proportional to $f_\pi$. Other
processes such as $\nu e\rightarrow\phi $ are also possible and their
contribution is proportional to the electron/positron
density. Further, since there are two relativistic particles in the
initial state (typical electrons have momenta $p\simeq 3 T \gg m_e$)
the kinematic constraints for producing a meson final state are
restrictive. Consequently, the contribution of these reactions is
found to be negligible.

The neutral current process involving the massless baryon
number Goldstone boson which we label as $H$ is identical to that in
the CFL phase. The contribution of reactions $\nu H\rightarrow \nu $
and $\nu \rightarrow H\nu $ to the neutrino opacity have been computed
previously \cite{Reddy:2002xc} and are given by
\begin{eqnarray}
\frac{1}{\lambda_{\nu\rightarrow H \nu}(E_\nu)}&=&
~\frac{256}{45\pi}~\left[\frac{v(1-v)^2(1+\frac{v}{4})}{(1+v)^2}
\right]~G_F^2~f_H^2 ~~E_{\nu}^3 \\
\frac{1}{\lambda_{\nu H \rightarrow \nu}(E_\nu)}&=&
\frac{128}{3\pi}\left[\frac{v~(1+v)^2}{(1-v)}\right]
\left[g_2(\gamma)+
\frac{2v~g_3(\gamma)}{(1-v)}-\frac{(1+v)~g_4(\gamma)
}{(1-v)}\right]
G_F^2~f_H^2~E_\nu^3
\end{eqnarray}
\ni where $\gamma=2vE_\nu/(1-v)T$ and the integrals $g_n(\gamma)$ are
defined by the relation $g_n(\gamma) = \int_{0}^{1}
dx~x^n/(\exp{(\gamma x)}-1)$. The decay constant for the $H$
mode has been derived earlier and is given by $f^2_H=3 \mu^2/(8
\pi^2)$. As we shall find below these reactions continue to be the
dominant source of opacity in the CFLK$^0$ phase.

To begin, we consider processes involving the $K^+$ Goldstone mode
since the $K^+$ is lightest charged mode. The leading
order processes are $\nu_e \rightarrow e^- K^+ $ and $\bar{\nu}_e 
K^+ \rightarrow e^+$.  The matrix element for these processes is given
by
\begin{eqnarray}
A_{K^+} &=& G ~f_\pi ~\sin{\tilde{\alpha}} ~\tilde{p}_\mu
\bar{e}(k_2)\gamma^\mu (1-\gamma_5)\nu (k_1) \,, \\
\label{amplitude}
{\rm where} \quad \tilde{p}^\mu &=& \left( E_{K^+} + X t_K,
v^2\vec{p}\right)\,, \quad P^\mu= \left(E_{\phi},\vec{p}\right) \,, \\
{\rm and} \quad t_{K} &=&
\left[\frac{\cos{\theta}\sin{\alpha}-\imag\sin{\theta}\cos{\alpha}}
{\sin{\tilde{\alpha}}}\right]~\cos{\frac{\theta}{2}}
\label{tildemom}
\end{eqnarray}
Here $\alpha$ is the vacuum Cabibo angle, $\theta$ is the condensate
amplitude and $\tilde{\alpha}$ (see Eq.~\ref{newcabibo}) is the
effective Cabibo angle in the $K^0$ phase. As discussed earlier, since
the $K^0$ condensate breaks $U(1)_Y$, it modifies the Cabibo
suppression of charged current reactions involving kaons. When the
condensate amplitude is large corresponding to $\cos{\theta} \simeq
0$, the rate of charged current weak reactions involving kaons becomes
independent of the $\alpha$ and are not Cabibo suppressed.

The opacity or inverse mean free path is defined to be the cross
section per unit volume. For the process $\nu_e \rightarrow e^- K^+ $
this is given by
\begin{eqnarray}
\frac{1}{\lambda_{\nu_e \rightarrow e^- K^+}} &=& \frac{1}{2
E_1}\int\frac{d^3p}{(2\pi )^32{\cal E}_{K^+}}\int\frac{d^3k_2}{(2\pi )^32E_2}
\sum_{spin}|A|^2(2\pi )^4 (1-f_{e^-} (E_2)) \delta (k_1-P-k_2) \,, \\
{\rm where}  \,, \nonumber \\
{\cal E}_{K^+}& =& \sqrt{v^2 p^2 + Y^2 + \delta m^2} \,,
\label{emit}
\end{eqnarray}
Note that $f_{e^-}(E_2)=(1+\exp((E_2-\mu_e)/T))^{-1}$ is the Fermi -
Dirac distribution function for the electrons and the factor
$(1-f_{e^-}(E_2))$ accounts for Pauli blocking of the final state
electron.  Since $K^+$ has a finite mass, neutrinos must have a
threshold energy to radiate them. Energy and momentum conservation
restrict the initial neutrino energy to have an energy $E_1 \ge E_{\rm
th}$ given by
\begin{eqnarray}
E_{\rm th} &=& \frac{\sqrt{y +\delta m^2 (1-v^2)}-y} {(1-v^2)} \, \nonumber \\
E_{\rm th} &\simeq& \tilde{m}_{K^+} \qquad {\rm for}\quad y \ll \delta m^2  \,.
\end{eqnarray}
Similarly, the opacity for the process $\bar{\nu}_e(k_1) K^+ (P)
\rightarrow e^+(k_2)$ is given by
\begin{equation}
\frac{1}{\lambda_{\bar{\nu}_e K^+ \rightarrow e^+ }} = \frac{1}{2 E_1}
\int\frac{d^3p}{(2\pi )^3 2{\cal
E}_K^+}f(E_{\phi})\int\frac{d^3k_2}{(2\pi )^32E_2}
\sum_{spin}|A|^2(2\pi )^4 (1-f_{e^+}(E_2)) \delta (k_1+P-k_2)
\label{absorb}
\end{equation}
where $f_B(E_{K^+})$ is a Bose distribution function for $K^+$ meson
and $f_{e^+}(E_2)= (1+\exp((E_2+\mu_e)/T))^{-1}$ is the Fermi - Dirac
distribution function for the electrons .

Reactions involving other Goldstone bosons make relatively small
contributions to the neutrino mean free path because they are
heavy compared to the $K^+$.  Nonetheless, it is straight
forward to calculate their contribution to the neutrino mean free
path.  The amplitude for reactions involving charged pions is obtained by
replacing $\sin{\tilde{\alpha}} \rightarrow \cos{\tilde{\alpha}}$ and
\begin{equation}
t_K \rightarrow t_{\pi} =
\left[\frac{\sin{\theta}\cos{\alpha}+\imag\cos{\theta}\sin{\alpha}}
{\cos{\tilde{\alpha}}}\right]~\sin{\frac{\theta}{2}}
\end{equation}
As noted earlier, for maximal condensation i.e., when $\cos{\theta}
\simeq 0$, the effective Cabibo angle
$|\cos{\tilde{\alpha}}|=1/\sqrt{2}$. Consequently, neutrino
reactions involving charged {\it pions} are mildly suppressed in the
$K^0$ phase.

Neutral current reactions involving $\pi^0$ and $\eta$ also
contribute to the neutrino opacity. Isospin breaking due to
kaon condensation results in a mixing between these states.  Like in
the vacuum, the $\pi^0$ and $\eta$ mix in the CFL phase due to isospin
breaking arises from $m_u \ne m_d$. However, since $m_d- m_u$ is still
small compared to other mass scales in the CFL phase, this mixing is
small (albeit moderately large compared to mixing in the vacuum). In
the CFLK$^0$ phase, we can compute the mixing from
Eq.\ref{mastereqn}. The mass matrix for the fields $\pi_3$ and $\pi_8$
corresponding to the $\pi_0$ and $\eta$ fields is given by 
\be
\Pi=\left[ \begin{array}{cc} \kappa_{33} & \kappa_{38}\\ \kappa_{83} &
\kappa_{88} \\
\end{array} \right] \,,
\label{pizeormass}
\ee
where
\begin{eqnarray}
\kappa_{33}&=&YZ+a\left[m_dm_s-m_um_s~\frac{Y}{X}+m_um_d\frac{Z}{X}\right]
\\
\kappa_{88}&=&3YZ+\frac{a}{3}~\left[m_d m_s (1+5\frac{Y}{X})
+m_um_s(1+5\frac{Z}{X})-m_um_d\right] \\
\kappa_{38}=\kappa_{83}&=&-\sqrt{3}YZ+\frac{a}{\sqrt{3}}~
\left[m_d m_s + m_u m_s \frac{Y}{X} +m_u m_d \frac{Z}{X} \right]
\end{eqnarray}
The unitary matrix
$U=\left[\begin{array}{cc}
\cos{\psi} & -\sin{\psi}\\
\sin{\psi} & \cos{\psi} \\
\end{array} \right]
$
that diagonalizes $\Pi$ determines the masses and mixing. The masses $m_{\tilde{\pi}_0}$ and $m_{\tilde{\eta}}$ corresponding to the mass eigenstates $\tilde{\pi}_0$ and $\tilde{\eta}$ are given
\be
\left[\begin{array}{cc}
m_{\tilde{\pi}_0}^2 & 0\\
0 & m_{\tilde{\eta}}^2 \\
\end{array} \right]
=~U~\Pi~U^{-1}\,,
\ee
and the eigenstates are given by
\be
\left[ \begin{array}{c}
\tilde{\pi}_0 \\
\tilde{\eta} \\
\end{array} \right]
=~U~\left[ \begin{array}{c}
\pi_0 \\
\eta \\
\end{array} \right] \,.
\ee
For the numerical values of parameters chosen in this work, we find
that in the CFLK$^0$ phase the mixing angle $\psi \simeq 25^o$ (in
the CFL phase, $\psi_{\rm CFL}\simeq 15^o$).

The amplitude for the tree level process such as $\nu \rightarrow
\nu \tilde{\pi}_0$ and $\nu \rightarrow \nu \tilde{\eta}$ are given by
\begin{eqnarray}
A_{\tilde{\pi}_0/\tilde{\eta}}&=&
\frac{G_F~C_{{\tilde{\pi}_0}/\tilde{\eta}}}{\sqrt{2}} \quad \tilde{p}^{\mu}~
\bar{\nu}(k_2)\gamma_\mu(1-\gamma_5)\nu(k_1) \\
{\rm where}\quad \tilde{p}_\mu&=&  (E_{\tilde{\pi}_0/\tilde{\eta}},v^2 \vec{p})
\,,
\end{eqnarray}
and neutral current coupling constants for the mass eigenstates are given by
\be
\left[ \begin{array}{c}
C_{\tilde{\pi}_0} \quad 
C_{\tilde{\eta}} \\
\end{array} \right]
=\left[ \begin{array}{c}
C_{\pi_0} \quad
C_{\eta} \\
\end{array} \right]~
~U^{-1}\,, \ee where $C_{\pi_0}=1$ and $C_{\eta}=1/\sqrt{3}$ are the
weak (isospin) charges of the $\pi_0$ and $\eta$ respectively. Contribution to the neutrino mean free path from the emission and
absorption of the neutral $\tilde{\pi}_0$ and $\tilde{\eta}$ mesons
can be computed using the charged current results in Eq.\ref{emit} and
Eq.\ref{absorb}, respectively but with the following substitutions:
$|A|^2 \rightarrow |A_{\tilde{\pi}_0/\tilde{\eta}}|^2$, $f_e(E_2)
\rightarrow f_\nu(E_2)=1/(\exp(E_2/T)+1)$, and
${\cal E}_K \rightarrow E_{\tilde{\pi}_0/\tilde{\eta}}=
\sqrt{v^2p^2+m_{\tilde{\pi}_0/\tilde{\eta}}^2}$.

We present numerical results for neutrino mean free path in the
CFLK$^0$ phase at $\mu=400$ MeV and for two different temperatures
$T=5$ MeV and $T=15$ MeV. For $T \gsim 15$ MeV we expect that finite
temperature corrections to the meson dispersion relations would play a
role. As discussed earlier, a rough dimension estimate of the critical
temperature at which the $K^0$ condensate would melt is give by $T_c
\sim 30-50$  MeV.  The results we present employ the following numerical values: $m_u=3.75$ MeV, $m_d = 7.5$ MeV, $m_s = 150$ MeV, $\Delta = 100$ MeV and $\chi = 0$.  The neutrino mean free path for thermal neutrinos ($E_{\nu}=\pi T$) due to different charged and
neutral current processes involving chiral Goldstone modes are
given in the table below. The results indicate that
reactions involving the $K^+$ are important.
\begin{center}
\begin{tabular}{|c|c|c|} \hline
process & $\lambda$(T=5 MeV) & $\lambda$(T=15 MeV) \\ \hline
$\nu_e\rightarrow K^+ e^-$ & 682 m & 42 m \\ \hline $\bar{\nu}_e
K^+\rightarrow e^+$ & 848 m & 43 m \\ \hline $\nu_e \rightarrow \pi^+
e^-$ & $\infty$ & 81 m \\ \hline $\bar{\nu}_e \pi^+\rightarrow e^+$ & 8.3 km
& 77 m \\ \hline $\bar{\nu}_e \rightarrow K^- e^+$ & $\infty$ &
$\infty$ \\ \hline $\nu_e K^-\rightarrow e^-$ & \gsim 10 km & 1.2 km \\
\hline 
\end{tabular}
\begin{tabular}{|c|c|c|} \hline
process & $\lambda$(T=5 MeV) & $\lambda$(T=15 MeV) \\ \hline
$\bar{\nu}_e \rightarrow \pi^- e^+$ & $\infty$ & $\infty$ \\
\hline $\nu_e \pi^-\rightarrow e^-$ & \gsim 10 km & 200 m \\ \hline 
$\nu \rightarrow \tilde{\pi}^0 \nu $ & $\infty$ & 6.9 km \\ \hline
$\nu \tilde{\pi}^0 \rightarrow \nu $ & \gsim 10 km &  350 m  \\ \hline
$\nu \rightarrow \tilde{\eta} \nu $ & $\infty$ & 1.6 km \\ \hline
$\nu \tilde{\eta} \rightarrow \nu $ & \gsim 10 km &  146 m \\ \hline
\end{tabular}
\end{center}
\begin{figure}[]
\begin{center}
\includegraphics[width=.48\textwidth,height=0.48\textwidth]{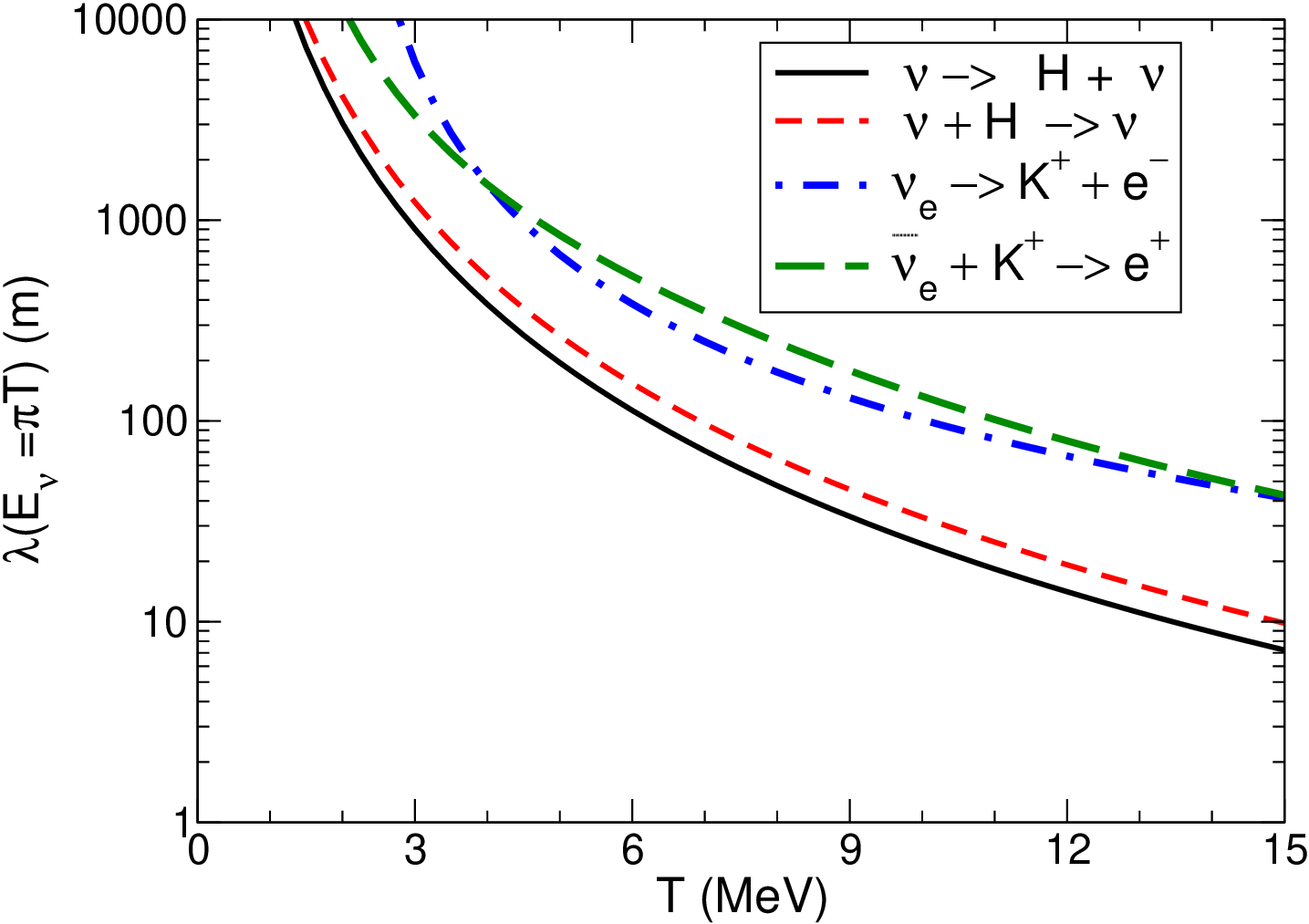}
\includegraphics[width=.48\textwidth,height=0.48\textwidth]{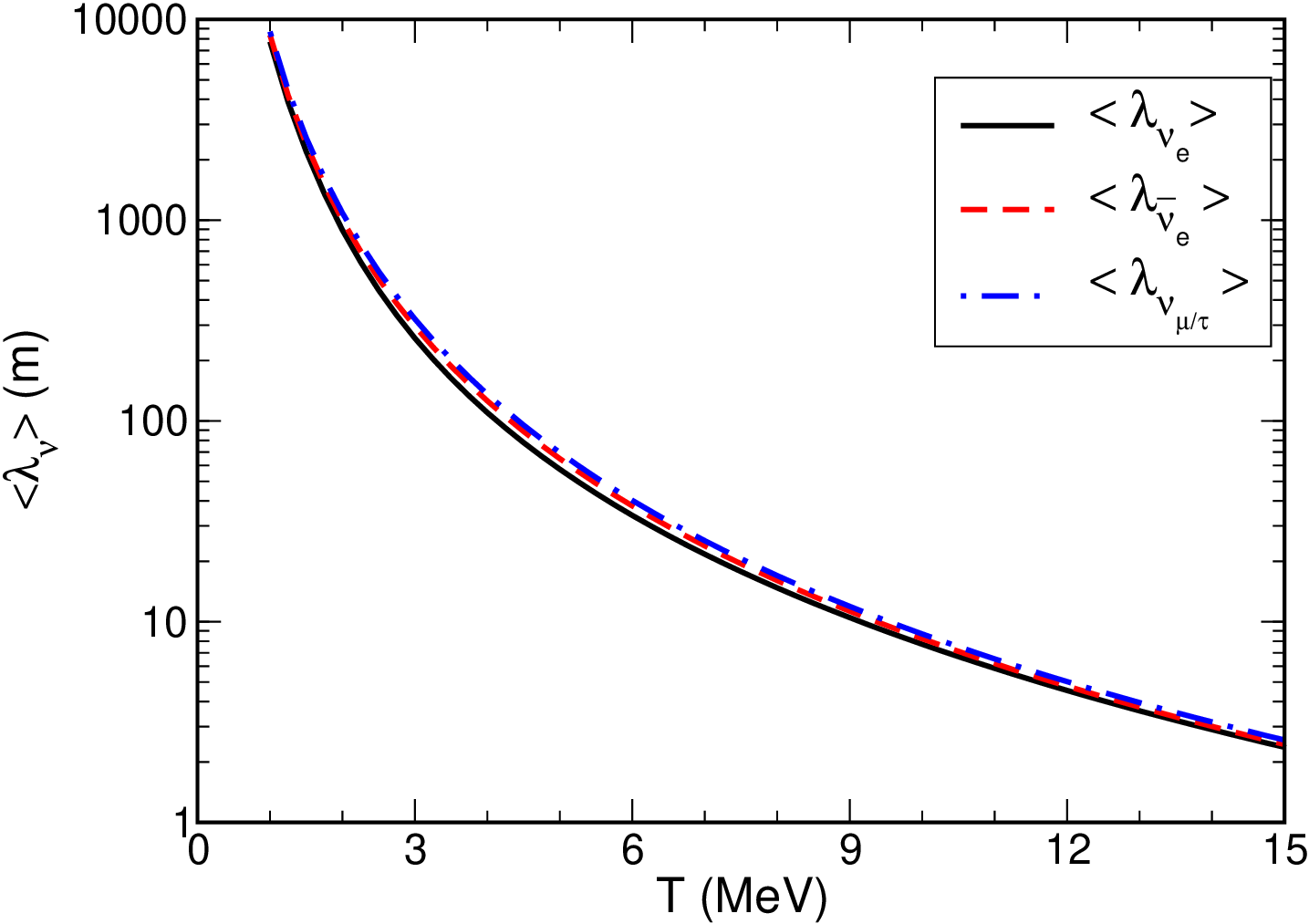}
\caption{Left Panel: Neutrino mean free path for a thermal neutrino
with energy $E_\nu = \pi T$ arising from dominant reactions in the
CFLK$^0$ phase. Right Panel: The average neutrino mean free path (defined in the text) in the CFLK$^0$ phase.}
\label{lambda}
\end{center}
\end{figure}
Since the $K^+$ meson is the lightest and hence the most abundant
flavor Goldstone mode reactions involving the $K^+$ dominate the
neutrino opacity.  In Fig.\ref{lambda} we show the neutrino mean free
path arising due to the dominant processes - those involving the $H$
and $K^+$ Goldstone modes.  The left panel shows the contribution of
individual reactions to the neutrino mean free path for thermal
neutrinos whose typical energy is $E_\nu=\pi T$. The right panel shows
the total Roseland averaged mean free paths for the different neutrino
species. The Roseland mean free path is defined as
\be \frac{1}{<\lambda_\nu>} =
\frac{1}{{\cal N}}~\int_0^\infty~dE_\nu~E_{\nu}^2
~\frac{1}{\lambda(E_\nu)}~f_\nu(E_\nu) \,, \ee
where ${\cal N}=\int_0^\infty~dE_\nu~E_{\nu}^2f_\nu(E_\nu)$ and
$f_\nu(E_\nu)=1/(1+\exp{(E_\nu/T)})$ is the distribution function for
thermal neutrinos. The neutral current process involving the H mode is
common to all neutrinos types, while the process $\nu_e \rightarrow
K^+ e^-$ is specific to the electron neutrinos and the reaction $\bar{\nu}_e K^+ \rightarrow e^+$ is specific to the anti-electron neutrinos.

\subsection{Neutrino emissivities}
We can compute the rate for these processes as outlined in
Ref.~\cite{Reddy:2002xc}.  The emissivity is defined to be the rate at
which energy is radiated in neutrinos per unit volume. To begin, we
consider the emissivity arising due to the decay of charged mesons
into electrons(positrons) and anti-neutrinos(neutrinos). The
contribution arising due to the decay of charge kaons can be computed
using
\begin{eqnarray}
\dot{\epsilon}_{K^{\pm} \rightarrow e^{\pm} \nu} &=&
\int \frac{d^3p~f_{K^\pm}(E_{K^{\pm}})}{(2\pi)^3~2 \CE_{K^{\pm}}}
\int \frac{d^3k_2~(1-f_{e^{\pm}})}
{(2\pi)^3~2 E_2} \nonumber \\
&\times& \int\frac{d^3 k_1}{(2\pi)^3~2 E_1}~ 
|A_{K^{\pm}\rightarrow e^{\pm} \nu}|^2
~(2 \pi)^4 ~ \delta^4(P-K_1-K_2) 
\label{emis}
\end{eqnarray}
where momenta $P=(E_{K^{\pm},\vec{p}}),K_2=(E_2,\vec{k}_2)$ and
$K_1=(E_1,\vec{k}_1)$ are the four momenta of the K$^{\pm}$, $e^\pm$
and neutrino respectively. The emissivity due to the decay charged
pions can also be computed from Eq.~\ref{emis} with appropriate
substitutions for the amplitude and the distribution function. The
inverse reactions such as $K^{\pm}+e^{\mp} \rightarrow \nu$ can also
be computed by replacing the factor $1-f_{e^\pm}$ by $f_{e^\mp}$
in the above equation. Emissivity due neutral current decays of 
$\tilde{\pi}^0$ and $\tilde{\eta}$ can be also be calculated using
Eq.~\ref{emis} by making the following substitutions:
$(1-f_{e^{\pm}})\rightarrow 1$, $A_{K^{\pm}\rightarrow e^{\pm} \nu}
\rightarrow A_{\tilde{\pi^0}/\tilde{\eta} \rightarrow e^{\pm} \nu}$
and $ \CE_{K^{\pm}} \rightarrow E_{\tilde{\pi^0}/\tilde{\eta}} =
\sqrt{v^2 p^2 + m_{\tilde{\pi^0}/\tilde{\eta}}^2}$. 

\begin{center}
\begin{tabular}{|c|c|c|} \hline
process & $\dot{\epsilon}$(T=5 MeV) & $\dot{\epsilon}$(T=15 MeV) \\ 
   &   ergs/cm$^3$/s & ergs/cm$^3$/s \\ \hline
$ K^+\rightarrow e^+ \nu_e$ & $7.4\times 10^{30}$ 
& $3.5\times 10^{31}$ \\ \hline
 $K^+ e^- \rightarrow\nu_e$ & $1.6\times 10^{34}$ 
& $6.9\times 10^{36}$ \\ \hline
$ \pi^+\rightarrow e^+ \nu_e$ & $1.4\times 10^{29}$ 
& $3.7\times 10^{30}$ \\ \hline
$\pi^+ e^- \rightarrow\nu_e$ & $1.1\times 10^{32}$ &
$1.6\times 10^{36}$ \\ \hline 
\end{tabular}
\begin{tabular}{|c|c|c|} \hline
process & $\dot{\epsilon}$(T=5 MeV) & $\dot{\epsilon}$(T=15 MeV) \\
   &   ergs/cm$^3$/s & ergs/cm$^3$/s \\ \hline
$ K^-\rightarrow e^- \bar{\nu}_e$ & $1.4\times 10^{32}$ & $1.1\times
10^{35}$ \\ \hline 
$\pi^- \rightarrow e^- \bar{\nu}_e$ &
$1.0\times 10^{33}$ & $2.6\times 10^{35}$ \\ \hline
$\tilde{\pi}_0\rightarrow\nu\bar{\nu} $ & $1.4\times 10^{35}$ &
$1.1\times 10^{36}$ \\ \hline $\tilde{\eta}\rightarrow\nu\bar{\nu} $ &
$1.2\times 10^{35}$ & $7.1\times 10^{35}$ \\ \hline
\end{tabular}
\end{center}
We present results for the emissivities arising due to the various
reactions in the CFLK$^0$ phase. Numerical results are presented in units
of ergs/cm$^3$/s and are for chosen values of physical parameters indicated
earlier. As expected at low temperature, the lightest mesons dominate
the emissivity. However, the decay of the $K^{+}$ meson is not an
efficient means of producing neutrinos. This is because the $K^{+}$
has a time-like four momentum only for momenta $p \lsim
m_{K^+}$. Thus, only a small fraction of the thermal $K^+$ modes can
decay. The inverse reaction $K^+ e^- \rightarrow \nu$, on the other
hand is more efficient, because in this case, a large fraction of the $K^+$ mesons with a space-like dispersion relation contribute. This is also true for reactions involving the $\pi^+$ mode. In contrast the more massive negatively charged modes have time-like dispersion relation over a wider range of momenta and for this reason their decays makes a relevant contribution to the total emissivity. Due to their larger masses these contributions become important at higher temperature.  The decay of neutral modes, which have intermediate mass, also make an important contribution to the total emissivity. 

\section{Caveats}
\label{caveats}
Prior to discussing how the results of this work will impact supernova
and neutron stars we briefly mention a few caveats to the results presented in this work. 

The size of the superconducting gap and the instanton induced $\langle \bar{q} q \rangle$ at these relatively low
densities is poorly known. If  $\langle \bar{q} q \rangle$ were large such that the $Tr [M\Sigma]$ contribution to the masses become relevant, then it will affect the mass spectrum of Goldstone modes and thereby directly alter the neutrino rates. In particular, the strange mesons would be too heavy and would decouple from the low energy response. 

In this work, we neglected the contribution of the $\eta'$ Goldstone
mode.  This was based on our naive expectation that the anomalous
contribution to its mass would be large. If this were not the case,
the $\eta'$ would be a very light state that would mix with the
$\pi^0$ and $\eta$ mesons \cite{Manuel:2000xt}. To study how a light
$\eta'$ Goldstone mode would affect weak interaction rates we have
computed the masses and mixing of neutral mesons. In doing so we
assumed that only $Tr[\tilde{M} \Sigma]$ operator contributes to the
masses. For the parameters used in this work we find that the
eigenstates $\tilde{\pi}^0, \tilde{\eta}$ and $\tilde{\eta}'$ have
masses given by $m_{\tilde{\pi}^0}=31.5$ MeV, $m_{\tilde{\eta}}=29.3$
MeV and $m_{\tilde{\eta}'}=3.7$ MeV, respectively. Their neutral
current couplings are given by $C_{\tilde{\pi}^0}=0.71$,
$C_{\tilde{\eta}}=0.29$ and $C_{\tilde{\eta}'}=0.96$, respectively.
Due to its small mass and large neutral current coupling the
$\tilde{\eta}'$ contribution to the neutrino mean free path due to the
processes $\nu + \tilde{\eta}' \rightarrow \nu$ and $\nu \rightarrow
\tilde{\eta}' + \nu $ are larger than similar reactions involving
$\tilde{\pi}^0$ and $\tilde{\eta}$.  At $T=5$ MeV: $ \lambda_{\nu +
\tilde{\eta}' \rightarrow \nu} = 2.9 $ km and $ \lambda_{\nu
\rightarrow \tilde{\eta}' + \nu} = 817 $ m; and at $T=15$ MeV:
$\lambda_{\nu + \tilde{\eta}' \rightarrow \nu} =101 $ m and
$\lambda_{\nu \rightarrow \tilde{\eta}' + \nu} = 177$ m. This is
comparable to the $K^+$ contribution to the opacity. The
emissivity due to the reaction $\eta' \rightarrow \nu \bar{\nu}$ is
small compared to those due the other neutral mesons. At $T=5$
MeV, $\dot{\epsilon}_{\tilde{\eta}' \rightarrow \nu \bar{\nu}} =3.5 \times 10^{30}$ ergs/cm$^3$/s and at $T=15$
MeV we find that it is $\dot{\epsilon}_ {\tilde{\eta}' \rightarrow \nu \bar{\nu}} =1.1 \times 10^{31}$ ergs/cm$^3$/s.

We have ignored the electromagnetic contribution to the mass of the
charged Goldstone modes. Earlier estimates indicate that this could be
of the order of a few MeV
\cite{Hong:2000ng,Manuel:2000xt,Kaplan:2001qk}. If $m_{\rm el} \gsim
\delta m$ it would alter our numerical results for the
rates involving the $K^+$ mode. However, we note that this can be
easily incorporated by substituting $\delta m^2 \rightarrow \delta m^2
+ m_{\rm el}^2$. Another source of concern is related to our use of
asymptotic expressions for the coefficients of the effective theory.

\section{Discussion}

We have studied neutrino reactions involving Goldstone bosons in
the kaon condensed phase of CFL quark matter. A light and electrically
charged $K^+$ mode is found to play an important role. The massless
neutral kaon mode does not play a role because it does not couple to
the neutrinos at leading order. In contrast, the massless mode
associated with the breaking of baryon number in the CFL phase is
unaffected by kaon condensation and continues to dominate the neutrino
opacity.  As in the CFL phase \cite{Reddy:2002xc}, space like
propagation of Goldstone modes allows for novel, Cerenkov-like,
processes in which neutrinos radiate or absorb Goldstone bosons as
they propagate in the medium. The amplitude for these leading order
processes is large since they are proportional $f_\pi \sim \mu$ . These
processes continue to dominate the neutrino opacity.

\subsection{Comparison between CFL and CFLK$^0$ phases:}
The mean free path of neutrinos in the CFL and CFLK$^0$ phases are
nearly equal. This is one of our main findings. This is because the
mean free path in the CFLK$^0$ phase continues to be dominated by the
neutral current process involving the U(1)$_B$ Goldstone mode. Thus,
the main conclusions drawn in Ref.\cite{Reddy:2002xc} regarding the
temporal aspects of the neutrino signal remain largely unchanged. 

Neutrino emission processes arise mainly due to reactions involving
massive pseudo-Goldstone modes. Reactions involving only one meson
dominate because these amplitudes are proportional $f_\pi \sim
\mu$. These include decays such as $K^{\pm} \rightarrow e^{\pm} \nu$,
$\tilde{\pi}^0 \rightarrow \nu \bar{\nu}$ and absorption reactions
such as $K^{\pm} e^{\mp} \rightarrow \nu$.  Reactions involving the
$K^+$, and the neutral $\tilde{\pi}^0$ and $\tilde{\eta}$ are dominant
at low temperature. We find the neutrino emissivity in the CFLK$^0$
phase to be roughly 1-2 orders of magnitude larger than in the CFL
phase for temperatures $T \lsim 15$ MeV. At $T=5$ MeV, the emissivity
in the CFL phase $\dot{\epsilon}_{\rm CFL} \simeq 5 \times 10^{33}$ ergs/
cm$^3$ /s, while in the CFLK$^0$ phase the emissivity
$\dot{\epsilon}_{{\rm CFLK}^0} \simeq 3 \times 10^{35}$ ergs/
cm$^3$/s. This enhancement in the CFLK$^0$ phase arises because:  (1) on average the Goldstone modes in the CFLK$^0$ phase
are lighter and (2) the Cabibo suppression of reactions involving
kaons in the CFL phase is greatly alleviated in the CFLK$^0$ phase.

\subsection{Comparisons between CFL/CFLK$^0$ and other phases}
Neutrino mean free path in the CFL and CFLK$^0$ phases are typically
larger than those nuclear phase. They are similar to those in the
unpaired quark matter under similar ambient conditions.  At $T=15$ MeV
and baryon density of $n_B=5~n_0$, the neutrino mean free path in the
CFL$K^0$ phases $\lambda_{{\rm CFLK}^0}\simeq 6$ m. The dominant
neutrino reaction in the nuclear phase is $\nu + n \rightarrow \nu +n$
results in a mean free path $\lambda_{\rm nuclear} \simeq 10$
cm. Under similar conditions the mean free path of neutrinos in
unpaired quark matter arising due to neutral current scattering off
quarks yields $\lambda_{\rm unpaired} \simeq 13$ m. We note that these
estimates ignore the possible role of strong interaction correlations
and account only for the Pauli blocking effects in these
reactions. Model calculations have shown that correlations in the
nuclear phase can greatly increase the neutrino mean free path (by a
factor of 3-5)
\cite{Horowitz:1991it,Burrows:1998cg,Reddy:1998hb}. Comparing the mean
free paths in these different phases we conclude that the mean free
path in unpaired quark and CFL/CFLK$^0$ phases are similar (same order
of magnitude) but are roughly about an order of magnitude larger than
in the nuclear phase.

The neutrino emissivity in the CFL \cite{Jaikumar:2002vg} and CFLK$^0$ phases are exponentially small compared to those in the nuclear phase for $T \ll 1$ MeV. This is because the meson masses are of the order of a few
MeV. The lightest excitation that can contribute to the emissivity is
the $K^+$ mode in the CFLK$^0$ phase. At $T=5$ MeV, the emissivity
in the CFL$K^0$ phase $\dot{\epsilon}_{{\rm CFLK}^0} \simeq 3 \times 10^{35}$ ergs/cm$^3$/s which is only an order of magnitude smaller than in the unpaired quark phase where $\dot{\epsilon}_{\rm unpaired} \simeq 3 \times 2 \times 10^{36}$ ergs/cm$^3$/s. At higher temperature the rates 
become more similar.  A general trend we see is that when the temperature become comparable to the mass of the lightest charged particle the rates in the CFL and CFLK$^0$ phases become comparable to those in the unpaired quark phase. 

\subsection{Astrophysical Implications}
The primary motivation for studying the neutrino mean free paths in
dense matter is core collapse supernova. The results obtained here
will have little impact on the long term cooling of neutron stars
which are characterized by temperatures $T \ll 1$ MeV. This is because
at these low temperatures, the neutrino mean free path is
large compared to the size of the neutron star. Massive Goldstone bosons with masses of order a few MeV are negligible and their contribution to the emissivity is exponentially small. In contrast, during
the first tens of seconds subsequent to the birth of the neutron star
neutrinos carry almost all (99 \%) of the Gravitational binding energy
($\sim 10^{53}$ ergs) stored inside the newly born hot "neutron" star,
with $T \sim 30-50$ MeV. The newly born star is also called the
proto-neutron star (PNS). The rate at which neutrinos diffuse and the
spectrum with which they decouple from the PNS can affect key aspects
of core collapse supernova - the explosion mechanism and r-process
nucleosynthesis.  For a galactic supernova, the several thousand
neutrino events predicted in detectors such as Super Kamiokande and
SNO will provide information about the propagation of neutrinos inside
the dense PNS.  In the discussion that follows we speculate on how
some the results obtained in this work might affect core collapse
supernova.

Since neutrino mean free path is dominated by neutral current
process in the CFL and CFLK$^0$ phases, the mean free path for all six
neutrino species are very nearly equal.  This is in contrast to what
is observed in the neutron-rich nuclear phase. Here, $\lambda_{\nu_e}
\lsim \lambda_{\bar{\nu}_e} \lsim \lambda_{\nu_{\mu / \tau}}$.  In the
nuclear phase the charged current reaction $\nu_e + n \rightarrow e^-
+ p$ dominates and the reaction $\bar{\nu}_e + p \rightarrow e^+ + n$
makes a smaller but relevant contribution to the mean free path while
$\mu / \tau$ neutrinos interact only through the neutral current
reactions.  As can be inferred from Fig. \ref{lambda}, the differences
between the mean free path of the different neutrino types is small in
the CFL$K^0$ phase because the charged current reactions involving the
$K^+$ mode makes only a modest contribution to the total mean free
path.  Nonetheless, it is interesting to note that the Roseland mean
free path show a trend that is similar to that of nuclear matter, i.e.
$<\lambda_{\nu_e}> \lsim <\lambda_{\bar{\nu}_e}> \lsim
<\lambda_{\nu_{\mu / \tau}}>$.

It is interesting to inquire if neutrinos decoupling from the quark
phase could have (relative) spectra that are different from those that
decouple from the nuclear phase. The spectra with which neutrinos
emerge from the PNS impact several observable aspects of supernova
such as the explosion mechanism, r-process nucleosynthesis and the
number of detected neutrinos in the terrestrial detectors. The
r-process in particular is sensitive to the relative spectra of the
electron and anti-electron type neutrinos since this determines the
neutron excess in the neutrino driven r-process wind in the supernova
(for a recent review see Ref.~\cite{Qian:2003ae}). If indeed, all six
neutrinos types emerge from the CFL phase with similar spectra it
could affect the neutron to proton ratio in the r-process wind. It is
premature to make definitive statements regarding how our findings
here would affect supernova observables since neutrino transport in
the PNS depends on several micro and macroscopic inputs. Nonetheless,
a robust finding of this work is that if neutrinos decouple from the
CFL or CFLK$^0$ phase the spectra of all six neutrino types will be
very similar.

The characteristic time that governs the rate of cooling by neutrinos
diffusion is given by $\tau_D = \bar{C}_V R^2/c\bar{\lambda}$, where
$\bar{C}_V$ is the average specific heat, $R$ is the size
of the diffusion region and $\bar{\lambda}$ is the typical neutrino
mean free path in the region \cite{Prakash:1997xs}. In nuclear and unpaired quark phases the specific heat per unit volume is large $c_{V} \sim \mu^2 T$, while in the CFL and CFLK$^0$ phases it is very small with $c_{V} \sim
T^3$. This is difference is likely to be the dominant effect that
distinguishes the cooling of PNS with CFL and CFLK$^0$ quark matter
from the other conventional scenarios. As noted earlier, the presence
of additional light degrees of freedom in the CFLK$^0$ leads to a
specific heat that is roughly twice as larger than that in the CFL
phase. This difference could also be a potentially important in distinguishing between the CFL and CFLK$^0$ phases.  

Incorporating the findings of this work in astrophysical
simulations of PNS evolution warrants much further work. This will allow one to make quantitative predictions for the supernova neutrino signal and r-process nucleosynthesis. A crucial, future, step in theoretical efforts to constrain novel high density phases with supernova observations. 


\vskip0.75in \centerline{\bf Acknowledgments} S. R. would like to
thank David Kaplan for useful discussions. We would like to thank Madappa Prakash and Thomas Schafer for reading the manuscript and for their  comments and suggestions  The research of S. R is
supported by the Department of Energy under contract W-7405-ENG-36.
The research of M.S. was supported by the Polish State Committee for
Scientific Research, (KBN) grant no.  2 P03B 09322. 
\bibliography{cflk0} \bibliographystyle{h-physrev3.bst}
\end{document}